# A Hardware/Firmware-Based Switching Gate Multiplexing Method for Pulse Mode Radiation Detectors

M. Rahman*, J. Mattingly

*Department of Nuclear Engineering, North Carolina State University, Raleigh, NC 27695, United States of America*

Abstract

We present a hardware/firmware-based switching gate multiplexing method for pulse mode radiation detectors that can combine many detector signals into two readout channels. One readout channel passes the signal of the multiplexed detector that "fired" first, and the other channel provides a variable-width logic pulse, i.e., a pulse width modulation (PWM) signal, that identifies the active detector. The multiplexed output pulse is produced by passing the first active detector's signal to a fan-in circuit by gating on the corresponding channel for a fixed duration while blocking all other detector signals. It does this using individual analog switches for all the detector signals. Each switch is controlled by a fixed width logic pulse that is triggered by the arrival of the first active detector pulse. Both the fixed width logic pulse and the PWM signal are generated using a field-programmable gate array (FPGA). To demonstrate the proposed multiplexing method, a prototype four-channel multiplexer was developed for use with four NaI(Tl) detectors. The performance of the multiplexer was evaluated in terms of its ability to retain energy resolution, timing resolution, and original pulse shape. The proposed multiplexing method showed very little degradation in energy resolution and timing resolution or alteration of pulse shape. The switching gate feature of the proposed method enables the multiplexer output to have very low noise contribution from the inactive channels. This multiplexing technique also has the unique capability of isolating and recovering the first active detector's output pulse in cases where there is overlap between pulses from different detectors in a single digitized record. These features make the proposed hardware/firmware-based switching gate multiplexing method very promising for application to large radiation detector networks.

## 1. Introduction

Multiplexing is a technique that combines multiple signals into a smaller number of channels. It is sometimes used to reduce the load on the data acquisition system of large radiation detector networks. Large radiation detector networks are used, for example, in fundamental physics experiments, neutron and gamma imaging systems, and medical imaging systems such as positron emission tomography (PET). If the data acquisition (DAQ) system employs digitizers, then this reduction in input load will ultimately reduce the number of analog-to-digital converters (ADCs) and the amount of memory required by the DAQ system, which would also reduce the total cost and power consumption.

Several multiplexing methods have been previously developed for radiation detection. Among them, charge division multiplexing [1, 2, 3, 4, 5], frequency domain multiplexing (FDM) [6, 7, 8, 9, 10], time modulation-based multiplexing [11, 12], and digital multiplexing [13, 14, 15] are the most frequently used methods. However, when used with large radiation detector networks, most of them exhibit degradation of energy resolution and/or timing resolution generally resulting from noise summation from inactive detectors.

In this paper, we present a hardware/firmware-based multiplexing method for pulse mode radiation detectors where a signal-driven logic-based algorithm, implemented on a field-programmable gate array (FPGA), controls analog switches to suppress noise from the inactive detectors. The multiplexer uses N individual analog switches for N detector signals and produces two output signals. One signal contains the multiplexed detector signal and the other contains a variable-width logic pulse, i.e., a pulse width modulation (PWM) signal that encodes the identity of the active detector. Because the proposed multiplexer blocks the inactive detector signals, noise from those detectors is not superimposed on the signal from the active detector. The new method is intended for applications in high channel density and low count rate measurements where the probability of coincidence between multiplexed detectors is low. The method minimally alters the original pulse shape, and it exhibits little degradation in both energy and timing resolution. It has the ability to isolate coincident pulses and recover the first active detector's output pulse. It can also detect corrupted records caused by coincident pulses from different detectors. These are significant improvements compared to existing multiplexing methods.

---

*Corresponding author
mrahman3@ncsu.edu (M. Rahman)

## 2. Proposed Method

In the proposed hardware/firmware-based switching gate multiplexing method, N pulse mode radiation detector output signals can be multiplexed into two readout channels. The readout channels will digitize two signals that contain the multiplexed detector pulse and a variable-width logic pulse that identifies the detector number. A diagram illustrating the multiplexing technique is shown in figure 1. Each pulse from a detector is passed through its own inverting amplifier[1]. The inverted signal is then transmitted through two paths: one path leads to a leading-edge discriminator (LED), and the other path leads to an adjustable signal delay unit[2]. The output of each LED is transmitted to an FPGA. On the FPGA board, three logic units were implemented: a Switch Trigger Logic unit, Detector Identification Logic unit, and an ADC Trigger Logic unit. Whenever a detector fires, a logic pulse from the corresponding LED is fed to an input of the FPGA board, and in response, the Switch Trigger Logic unit produces a fixed width logic pulse (termed the control pulse) at the corresponding output channel of the FPGA board. The Switch Trigger Logic unit has the ability to isolate overlapping events. If multiple detectors fire nearly simultaneously, it produces the control pulse only at the output channel that corresponds to the detector that fired first.

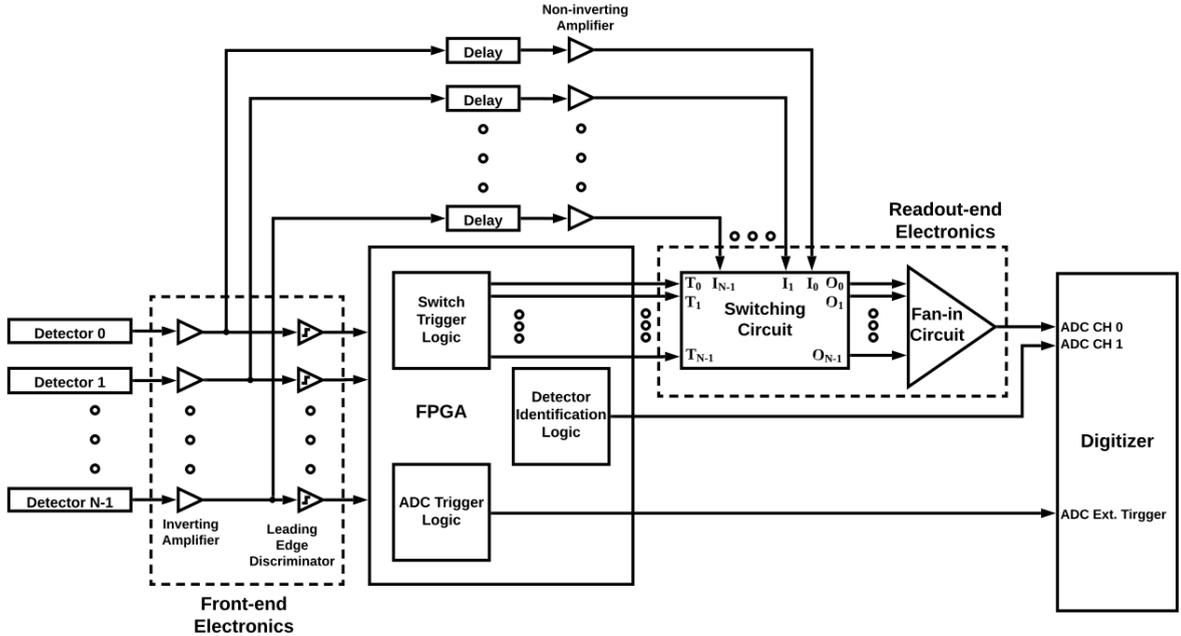

**Figure 1:** Diagram of the hardware/firmware-based switching gate multiplexer.

The control signal from the Switch Trigger Logic unit of the FPGA is fed to the $T_n$ input of the switching circuit. Simultaneously, the delayed detector pulse is fed to the $I_n$ input of the switching circuit; here, n = 0,1, 2, …, N-1. The switching circuit turns on and passes the input signal at $I_n$ to the output $O_n$ only when the $T_n$ input is in the "high" state. For an N channel multiplexer, there are N switches in the switching circuit, and only the switch that corresponds to the first active detector is activated. The output from the switching circuit is routed to a fan-in circuit which outputs the multiplexed signal. Switching off the N-1 inactive detector signals blocks noise from those signals. The switching circuit and the fan-in circuit form the readout-end electronics of this multiplexing scheme. The Detector Identification

---

[1] Anode pulses were required to be inverted because most of the comparator ICs (for leading edge discriminator) and analog switch ICs accept only positive polarity input signals.
[2] The delay unit is used to synchronize the detector pulse with the control pulse generated by the FPGA.

Logic unit, implemented on the FPGA, produces a delay-tunable pulse width modulation (PWM) signal. The width of the PWM pulse identifies the active detector for each event. The fan-in circuit output and the identification signal are transmitted to two separate ADC channels of the digitizer. The ADC Trigger Logic unit produces a delay-tunable logic pulse (termed the trigger pulse) that activates the ADC channels to record the output of the fan-in circuit and the corresponding PWM signal only when a detector fires.

## 3. Hardware and Firmware Implementation
### 3.1. Circuit Implementation

A schematic of a prototype four-channel multiplexer circuit is shown in figure 2. The circuit can be divided into three parts: (a) the front-end electronics consisting of inverting amplifiers and leading-edge discriminators (LEDs); (b) an attenuated signal amplifier and copier (ASAC) module consisting of two non-inverting amplifiers for each channel; and (c) the readout-end electronics consisting of a switching circuit and a fan-in circuit. In this schematic, the circuit components are shown only for channel number 0 which carries the signal from detector number 0.

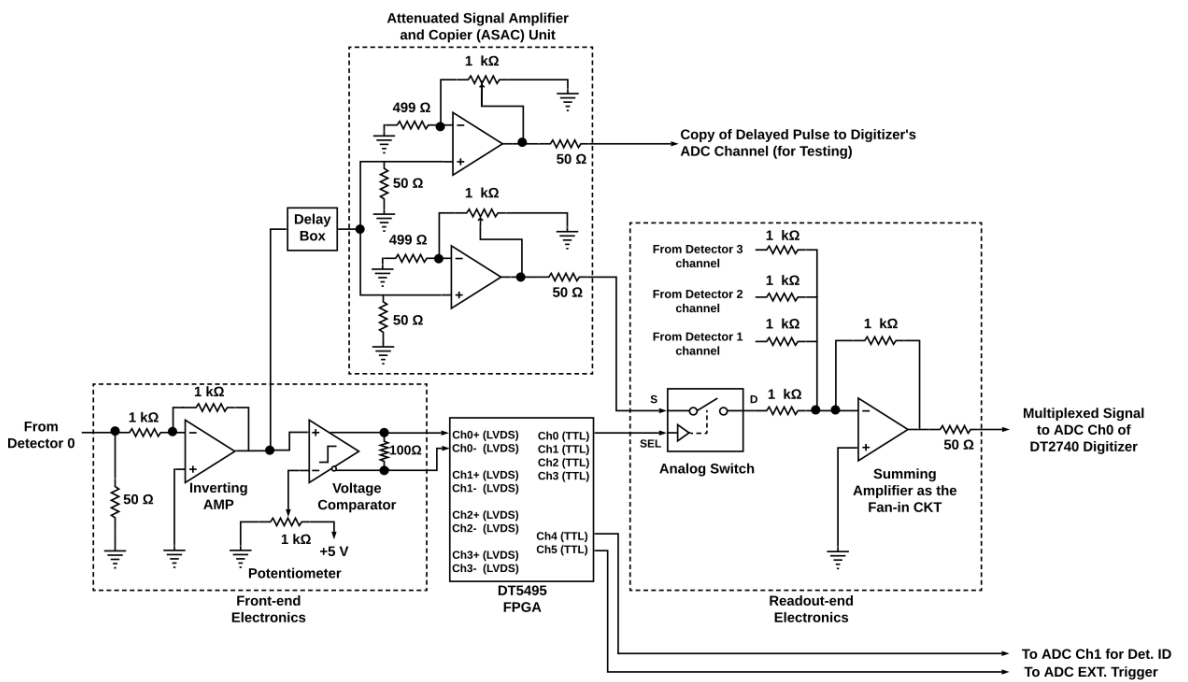

**Figure 2:** Schematic of the prototype 4-channel multiplexer circuit.

In the front-end electronics, the inverting amplifier was implemented using OPA820, a low-noise, voltage feedback operational amplifier [16]. The gain was set to -1 to invert the anode pulse without amplifying it. The leading-edge discriminator was implemented using TLV3604, a high-speed (800 ps), rail-to-rail input, voltage comparator with Low Voltage Differential Signal (LVDS) outputs [17]. For each channel, the non-inverting and inverting LVDS outputs from the TLV3604 are transmitted to one of the LVDS input channels (Ch0+ and Ch0- ports, respectively, for channel 0) of the FPGA.

The inverted detector pulse was delayed using a Model DB463 delay box [18]. The delay unit is followed by an attenuated signal amplifier and copier (ASAC) unit containing two non-inverting amplifiers implemented using OPA820 ICs. The gain can be adjusted by tuning a 1 kΩ potentiometer to compensate for signal attenuation by the delay line. These two amplifiers transmit two copies of the input signal through two paths. One path takes the delayed signal to the switching circuit while the other path can be connected to an ADC channel of the digitizer to simultaneously digitize the original pulse[3]; this signal was used solely for testing and evaluation described later in this paper.

---
[3]The delayed inverted detector pulse from the ASAC unit is subsequently referred to as the **original pulse**.

For the four-channel prototype, the switching circuit consists of four analog switches. Each switch was implemented using TMUX1101, a precision CMOS single-pole, single-throw (SPST) switch [19]. The FPGA board was programmed to produce a control pulse in response to an LVDS pulse from the LED. When a detector fires, this control pulse activates the analog switch for a fixed duration (equal to the width of the TTL logic pulse) to pass the delayed pulse to the fan-in circuit. An inverting summing amplifier, which also used OPA820, was implemented as the fan-in circuit in this prototype.

Four different printed circuit boards (PCBs) were fabricated to implement the front-end electronics, attenuated signal amplifier and copier (ASAC) module, switching circuit, and the fan-in circuit, respectively (figure 3). The front-end electronics were fabricated on a 2-layer PCB, while the other three boards were fabricated on a 4-layer PCB. The PCBs are based on flame retardant-4 (FR-4) substrate and the traces are impedance controlled (50Ω). Each PCB, except the fan-in circuit, consists of four independent input and output channels marked as Ch0, Ch1, Ch2 and Ch3. The fan-in circuit PCB has four input channels (also marked as Ch0, Ch1, Ch2 and Ch3) and only one output channel.

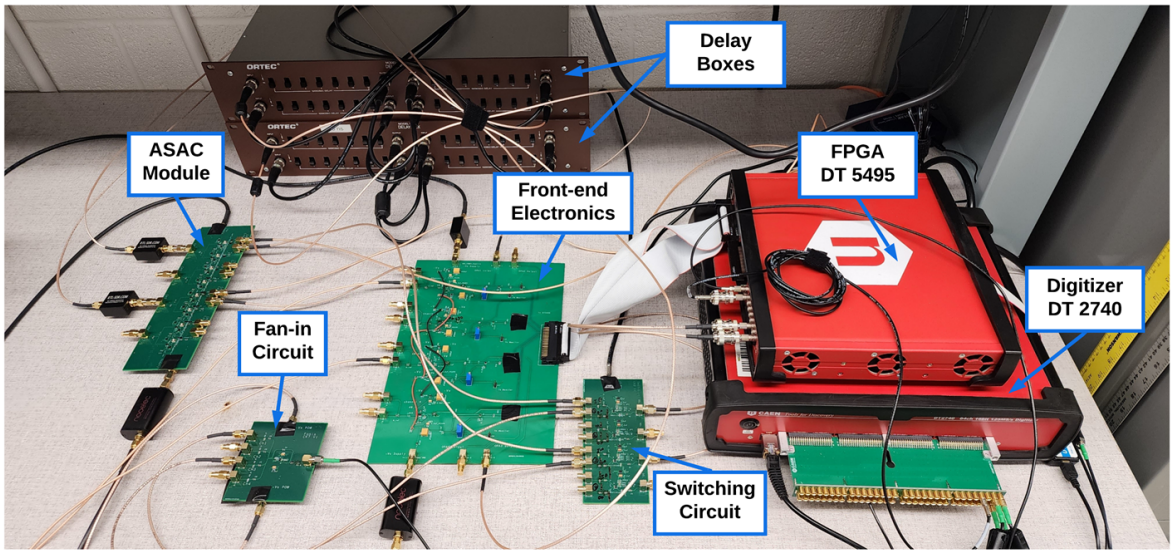

**Figure 3:** Prototype of the multiplexer with two of the four channels connected.

### 3.2. Logic Implementation

The functions to generate the control pulse, identification pulse, and trigger pulse were implemented on an FPGA board. The prototype multiplexer used DT5495, a programmable logic unit (PLU) [20] based on the Altera Cyclone V GX C4. The FPGA is directly interfaced to the front panel input-output ports and to an onboard Gate and Delay Generator. It was programmed using Sci-Compiler software, a block diagram-based programming tool compatible with CAEN Open FPGA boards [21].

In the following designs, all the functional blocks that require clock signal inputs are synchronized with a single 200 MHz clock. For simplicity, the clock connections are not shown in the logic diagrams.

#### 3.2.1. Switch Trigger Logic Implementation

The logic diagram of the Switch Trigger Logic unit is shown in figure 4. Each LVDS input signal is passed through a Pulse Shaper unit which samples the input signal at the rising-edge of the clock signal and produces an output pulse with programmable width. The width is expressed in terms of the number of clock cycles, and the value is set on a register connected to the Pulse Width input of the Pulse Shaper unit. The purpose of using this Pulse Shaper unit is to generate programmable pulses with a fixed width greater than the input pulses. An example timing diagram of a Pulse Shaper is shown in figure 5(a). For a four-channel multiplexer, there are four Pulse Shaper units at the input stage of the Switch Trigger Logic unit, and all of them have their Pulse Width input set at 650 clock cycles (pulse width = 3250 ns). The outputs of the four Pulse Shaper units are transmitted through two paths: one path takes the four outputs to an OR gate, and the other path takes them to four different two-input AND gates. The output of the OR gate is transmitted to an Edge Detector unit which detects the rising edge of the input pulse and produces an output pulse with a programmable width. An example timing diagram of an Edge Detector is shown in figure 5(b).

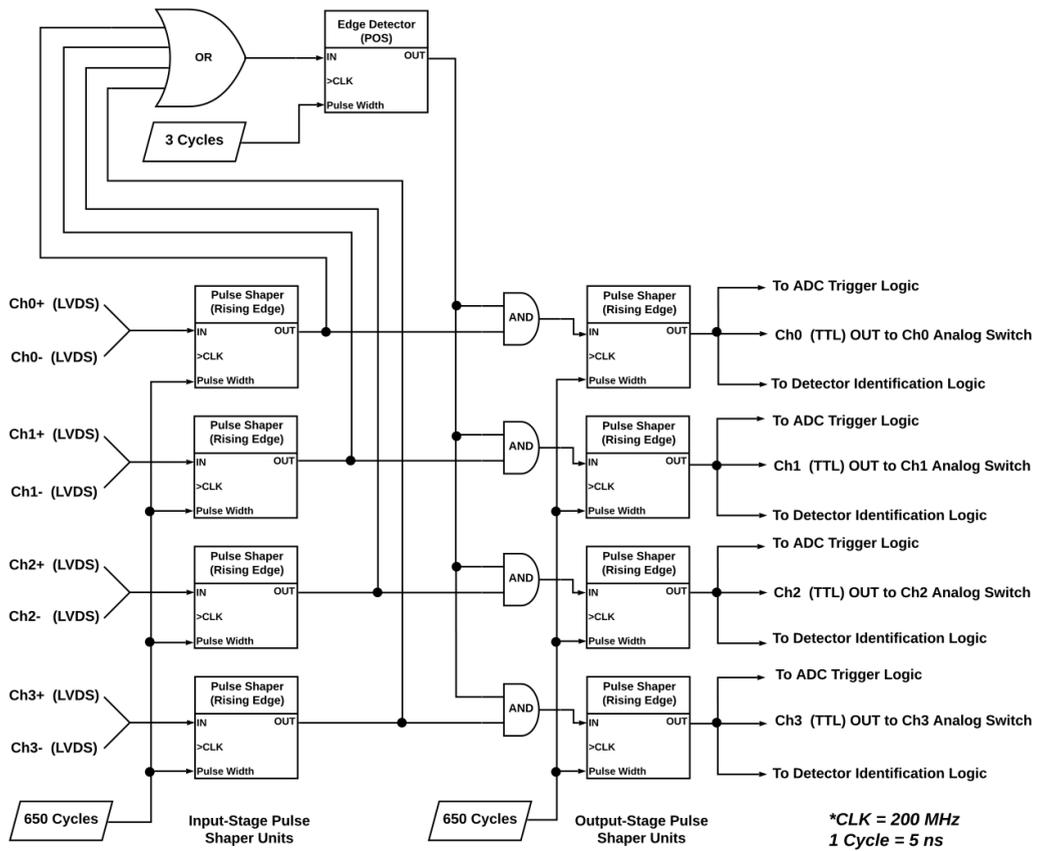

**Figure 4:** Logic diagram of the Switch Trigger Logic unit.

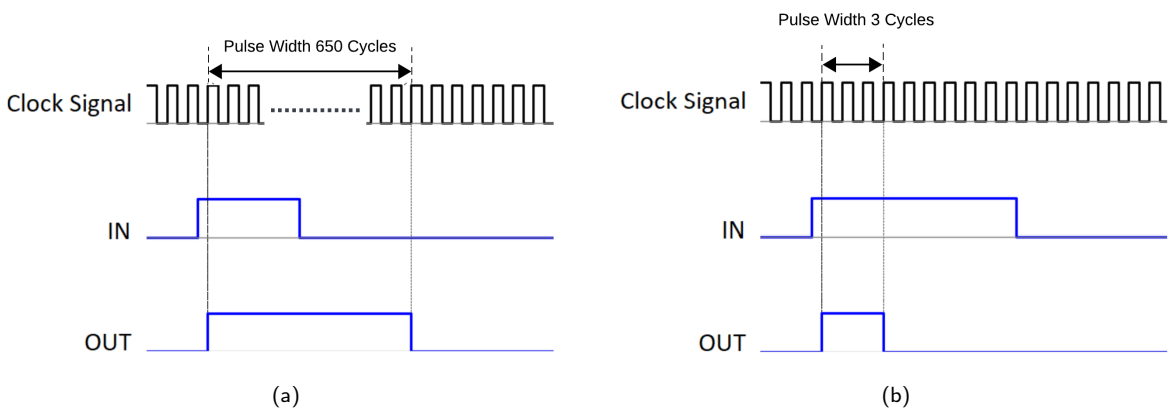

**Figure 5:** Timing diagrams for (a) Pulse Shaper and (b) Edge Detector.

The width of the Edge Detector is set at 3 clock cycles, so it generates a 15 ns wide output pulse whenever the output of the OR gate changes its state from LOW to HIGH. If two or more detectors fire within the fixed width time interval (3250 ns) (figure 6(b) and 6(c)), then the width of the OR gate output pulse will be greater than the fixed width (figure 6(d)). In response, the Edge Detector generates a 15 ns output pulse (figure 6(e)) that marks the arrival of the anode pulse from the detector that fired first. The output of the Edge Detector unit is transmitted to the second input of all four two-input AND gates. Among the four AND gates, only the one which is connected to the first active detector produces a 15 ns pulse (figure 6(f)).

The four AND gate outputs are applied to four output-stage Pulse Shaper units. The 15 ns narrow pulse produced by the AND gate, which corresponds to the first active detector, induces a corresponding Pulse Shaper unit to produce a fixed width pulse of 3250 ns (figure 6(h)) while the other Pulse Shaper units of the output stage produce LOW outputs (figure 6(i)). Here, the output signals of the four output-stage Pulse Shaper units function as the control pulses. They are transmitted to corresponding analog switches from the TTL output ports of DT5495. The same outputs are also passed to the ADC Trigger Logic unit and Detector Identification Logic unit. The input stage Pulse Shaper units, OR gate, Edge Detector unit, and the AND gates block nearly simultaneously firing (i.e., coincident) detector pulses from the fan-in circuit, and they ensure that only the first active detector's signal is transmitted to the fan-in circuit.

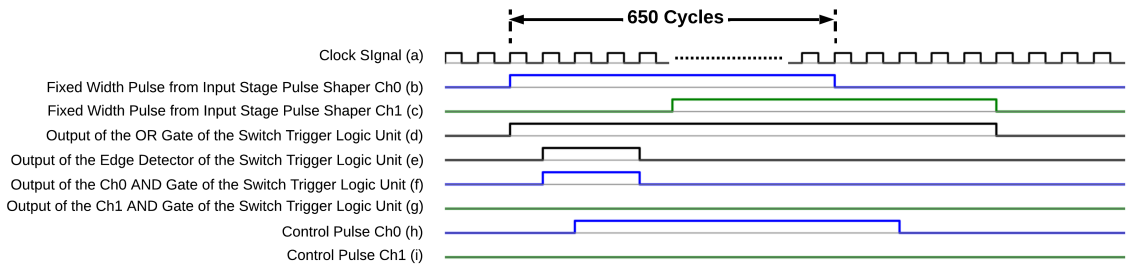

**Figure 6:** An example timing diagram of the Switch Trigger Logic unit using channel 0 and 1 only.

### 3.2.2. Detector Identification Logic Implementation

For the four-channel multiplexer, control pulses from the Switch Trigger Logic unit are applied to the Edge Detectors of the Detector Identification Logic unit (figure 7). The Pulse Width input of the four Edge Detectors, from channel 0 to channel 3, are set at 20 cycles, 40 cycles, 60 cycles, and 80 cycles, respectively. When a detector fires, the corresponding channel's Edge Detector generates a pulse with a preset width. This Pulse Width Modulation (PWM) signal encodes the identity of the active detector in its pulse width. The PWM signal outputs from all four channel Edge Detectors are fed to an XOR gate which functions as a 4-to-1 logic fan-in component. The output of the XOR gate (identification pulse) is passed through a Delay unit that delays it by the number of clock cycles set at its Delay input. This delay is used to synchronize the identification pulse with the corresponding multiplexed signal to simultaneously recover the detector pulse and identify the active detector. The delayed identification pulse is sent to one of the ADC channels of the digitizer.

In cases when multiple detectors fire nearly simultaneously, fixed width pulses from multiple input-stage Pulse Shapers of the Switch Trigger Logic unit may overlap within the width of the 15 ns output pulse from the Edge Detector of the Switch Trigger Logic unit. In such a case, the blocking mechanism implemented in the Switch Trigger Logic unit does not work, and the switches corresponding to all the active detectors are turned on. As a result, the multiplexed output from the fan-in circuit will be the summation of the coincident detector pulses. In the Detector Identification Logic unit, the fan-in implemented with an XOR gate acts as an anti-coincidence gate. It transmits the identification pulse with correct width only when there are no overlapping pulses within 15 ns pulse switch trigger (figure 8(a)). In the case of partial overlap, the XOR gate alters the identification pulse by producing multiple pulses instead of a single pulse (figure 8(b)). In the case of complete overlap, the XOR gate alters the identification pulse by shifting the pulse to the right relative to its normal position in the digitized record (figure 8(c)). By identifying altered identification pulses, events corresponding to summed coincident detector pulses can be removed during post-processing.

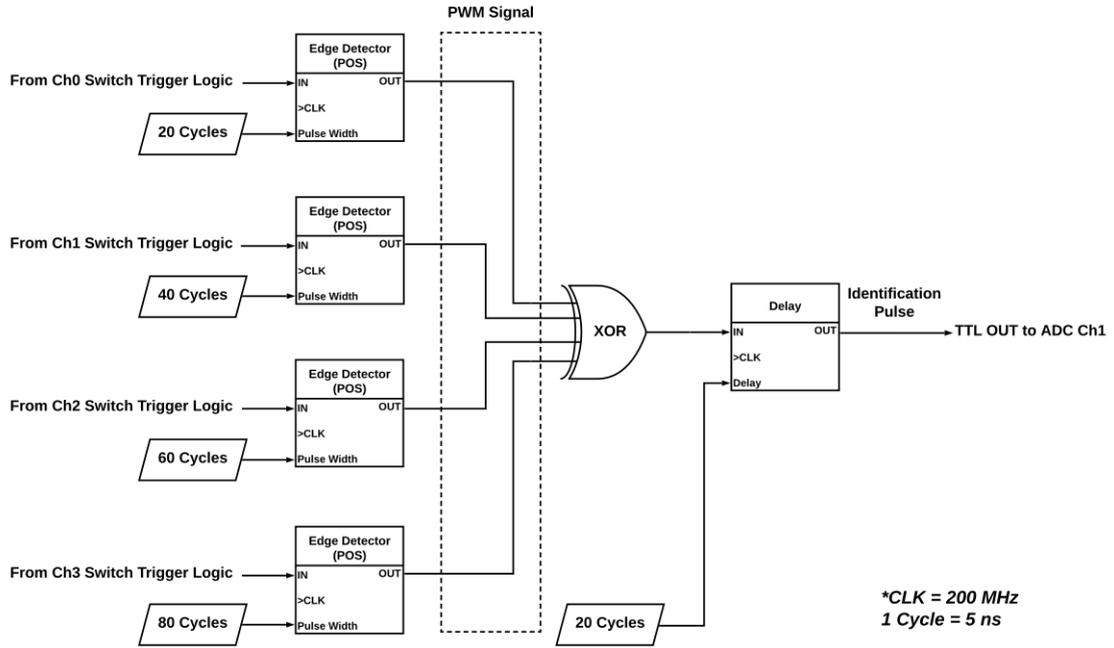

**Figure 7:** Logic diagram of the Detector Identification Logic unit.

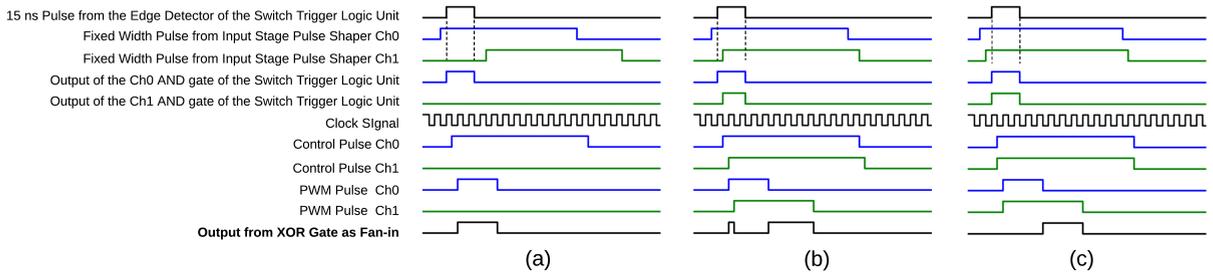

**Figure 8:** Timing diagrams for nearly simultaneous firings of detector 0 and detector 1 when an XOR gate is used as the fan-in. (a) No overlap, (b) partial overlap, and (c) complete overlap of fixed width pulses from input stage Pulse Shapers within the 15 ns pulse duration.

### 3.2.3. ADC Trigger Logic Implementation

Control pulses from the Switch Trigger Logic unit are transmitted to an OR gate of the ADC Trigger Logic unit (figure 9). This OR gate functions as a 4-to-1 logic fan-in component; the output of the gate is the trigger pulse. This trigger pulse is then passed through a Delay unit which synchronizes it with the corresponding multiplexed and identification pulses sent to the digitizer. The delayed trigger pulse is sent to the External Trigger input of the digitizer.

## 4. Setup for Testing and Evaluation

To evaluate the performance of the multiplexer prototype, the energy resolution, timing resolution and pulse shape parameters (rise-time constant and fall-time constant) acquired using the multiplexer were compared to measurements without the multiplexer. The setup for necessary measurements are described in this section.

### 4.1. Energy Resolution and Pulse Shape

Measurements were conducted to evaluate the performance of the proposed multiplexing scheme in terms of retaining energy resolution and original pulse shape. The measurement used four $2''\times4''\times16''$ NaI (Tl) scintillators each coupled to its own 3.5″-diameter 10-stage photomultiplier tube (PMT). A $^{137}$Cs source (approximately 15 $\mu$Ci) was used as the gamma source. First, the energy resolution of each detector was obtained using an analog pulse processing system. A Scionix Holland AMP100-E2 type preamplifier, ORTEC model 572A amplifier, and ORTEC model EASY-MCA-8k multi-channel analyzer (MCA) were used for this measurement. Energy resolution was measured from the full-width at half-maximum (FWHM) of the 662 keV photopeak. Energy resolution measured using this analog pulse processing system represents the best possible energy resolution that we could obtain, so it serves as a benchmark.

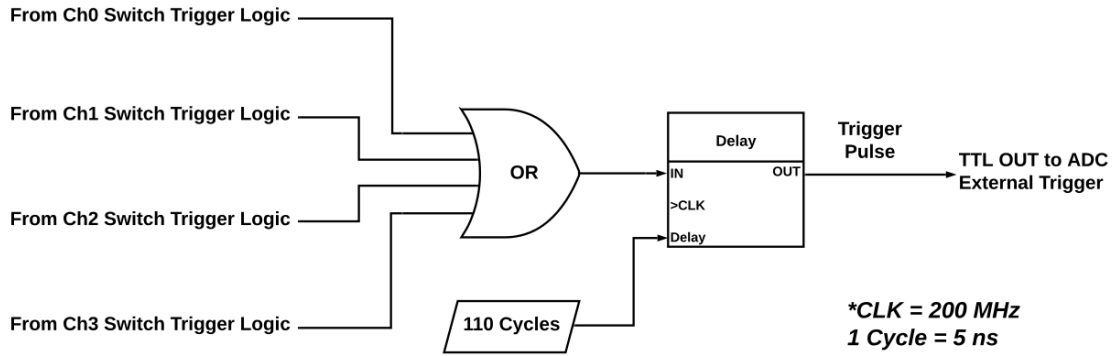

**Figure 9:** Logic diagram of the ADC Trigger Logic unit.

After measuring the energy resolution of the four detectors using an analog pulse processing system, four different multiplexer systems were set up using one, two, three, and then four detectors.

The instrumentation setup was the same as shown in figure 3. The detectors were placed parallel to each other on a bench, and the $^{137}$Cs source was placed in front of the front-face of the detectors.

The value of delay was set to 270 ns for all channels to synchronize the anode pulse with the corresponding switch control pulse. This delay provided sufficient baseline at the start of each recorded pulse to estimate and subtract the average baseline.

The settings specified in figures 4, 7, and 9 were applied to generate the control pulse, the identification pulse, and the ADC trigger pulse.

A DT2740 digitizer was used to simultaneously digitize and record the original pulses from the ASAC module and the multiplexed pulses from the fan-in circuit. DT2740 is a 16-bit, 64-channel digitizer with a sampling rate of 125 MS/s [22]. Digitization of each pulse started at the rising edge of the trigger pulse. The record length was set to 3040 ns to digitize the entire anode pulse and to match the width of the control pulse, which was set to 650 clock cycles or 3250 ns. The baseline of each pulse was removed by subtracting the mean amplitude of the first 5 samples (40 ns) of the pulse. Charge was obtained by trapezoidal integration of each baseline-subtracted pulse, and the energy spectrum was constructed from the histogram of charge. The energy resolution was estimated by fitting a Gaussian to the 662 keV photopeak.

### 4.2. Timing Resolution Measurement

To measure the performance of the proposed multiplexer in terms of timing resolution, two identical one-channel multiplexers were set up using two NaI(Tl) detectors and a 0.9 $\mu$Ci $^{22}$Na source. The positron emitted by $^{22}$Na $\beta^+$ decay annihilates with an electron to produce two 511 keV annihilation photons traveling in opposite directions. Two detectors were placed 30 cm apart, and the source was placed between them to detect the coincident photons. The connections were identical to the energy resolution measurement, except that in this case, two one-channel multiplexers were set up by using two independent channels of the four-channel prototype. Each channel had an independent fan-in circuit. Detector pulses from the two analog switches were fed to the two fan-in circuits. The logic parameter setup was also similar to that of the energy resolution measurement, except that in this case, the identification pulse was unused. A coincidence detection logic was set up in the FPGA; it used a window of 50 ns (10 clock cycles). By applying constant fraction discrimination (CFD) at the rising edge of the recorded pulses, the time-of-arrival (TOA) of the original pulses and the multiplexed pulses for both detectors were measured. From the measured TOA values, the difference between the TOA for the two coincident detections were calculated. For both original and multiplexed cases, the distribution of the TOA difference was constructed, and the timing resolution was estimated by fitting a Gaussian to the peak of the coincidence histogram.

## 5. Results and Discussion

### 5.1. Basic Outputs

Oscilloscope traces for an event from a two-channel multiplexer using physical channels 0 and 2 is shown in figure 10. In this event, only detector 0 fired and the control pulse at the channel 0 analog switch was activated. As a result, the fan-in circuit received the detector 0 delayed pulse at its input and inverted the pulse at its output. A shift in the DC level of the fan-in output is visible when the analog switch is turned on. This shift in DC level corresponds to the summation of the output offset voltages from the op-amps and the analog switch of channel 0. Spikes in the fan-in output can be observed at both ends of the transition of the switch. These spikes can be attributed to the charge injection phenomenon present in MOS-based analog switches or transmission gates [23]. Stray capacitance accompanying the NMOS and PMOS transistors, that make up the MOS based analog switch, is the major reason for charge injection. To reduce the charge injection, four ceramic capacitors (two 0.1 $\mu$F and two 1 $\mu$F) and two tantalum capacitors (one 2.2 $\mu$F and one 10 $\mu$F) were used to decouple the power pins of each analog switch. This technique reduced the amplitude of the spikes but it was still comparable to lower amplitude detector pulses.

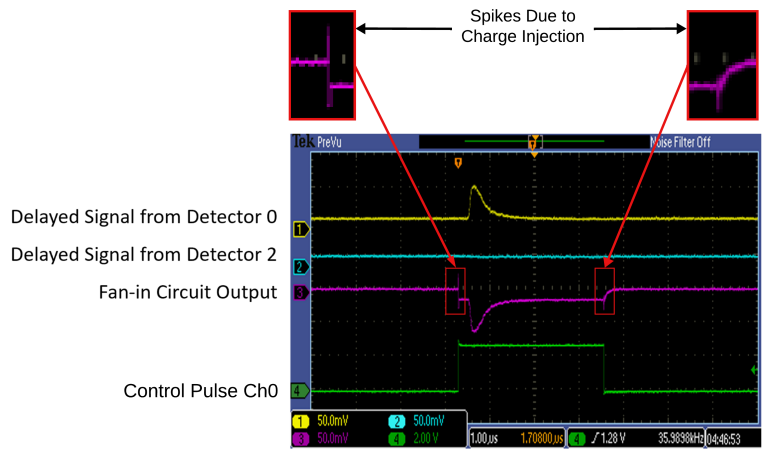

**Figure 10:** Oscilloscope traces for an event from a two-channel multiplexer.

To avoid spikes in the digitized record, the synchronization of the ADC trigger pulse with the fan-in started just after the first spike and ended just before the second spike. Two such sample events are shown in figures 11 and 12.

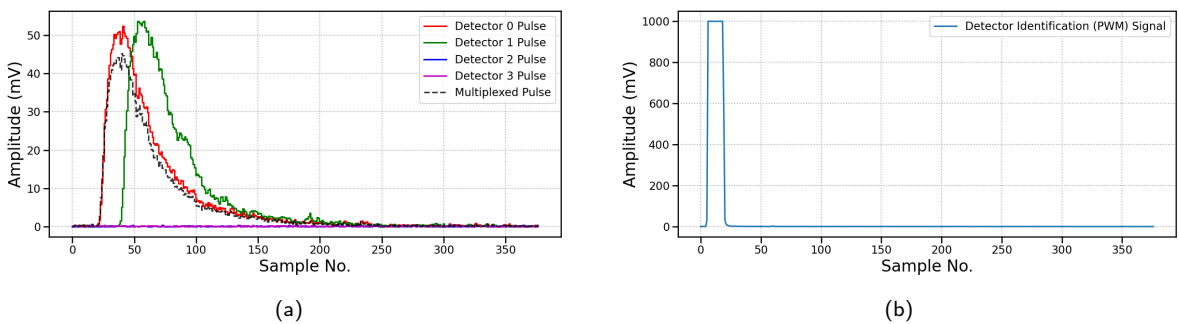

**Figure 11:** A sample event recorded with digitizer where detector 0 fired before detector 1. (a) Original pulses from all four detectors and the multiplexed pulse. (b) Identification pulse indicating detector 0 fired first.

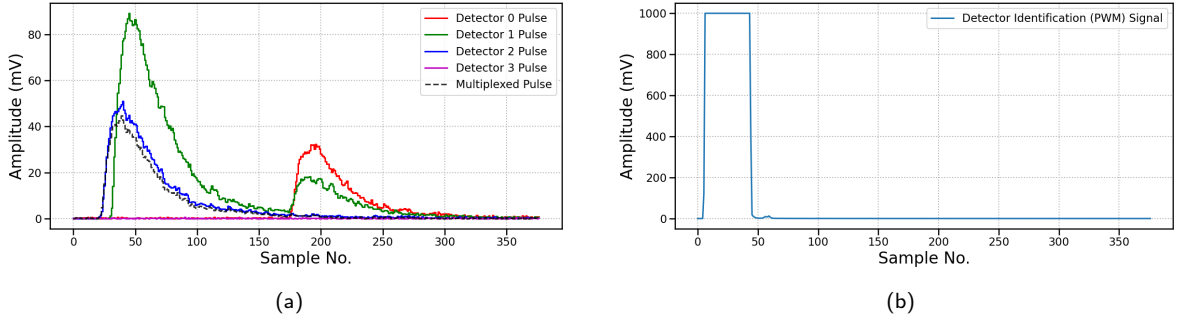

**Figure 12:** A sample event recorded with digitizer where detector 2 fired before detector 0 and 1. (a) Original pulses from all four detectors and the multiplexed pulse. (b) Identification pulse indicating detector 2 fired first.

Figures 11 and 12 also demonstrate the effectiveness of the blocking mechanism implemented in the proposed multiplexing technique. Although multiple detectors fired closely spaced in time, pulses of the detectors did not overlap in the multiplexed signal in either case. For the event in figure 11, the multiplexed pulse was the first arriving pulse from detector 0. For the event in figure 12, the multiplexed pulse was the first arriving pulse from detector 2. Also, in both cases, the identification pulse was able to identify the first active detector (figures 11(b) and 12(b)).

The effectiveness of using an XOR gate as the fan-in in the Detector Identification Logic unit is shown in figures 13 and 14. In both cases, detector 0 and detector 1 fired simultaneously, and the fixed width pulses overlapped within the 15 ns pulse width of the Edge Detector output. This resulted in the summation of the two detector pulses at the multiplexer output. The XOR gate tagged these kinds of events by altering the identification pulse as shown in figures 13(b) and 14(b).

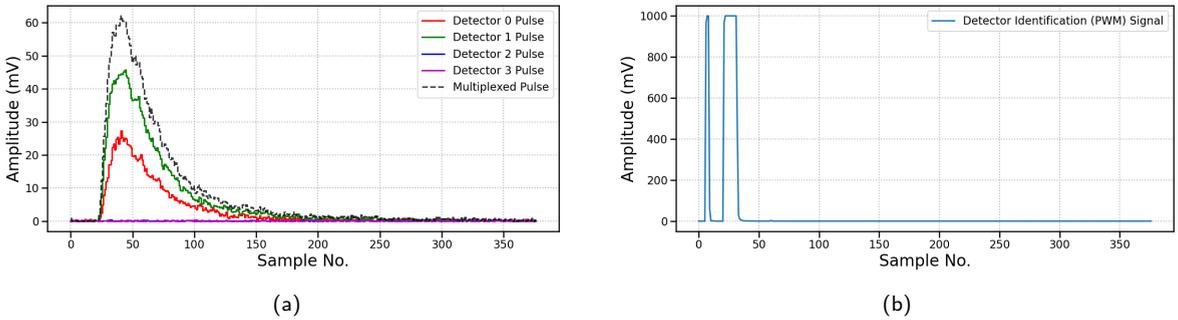

**Figure 13:** Demonstration of the XOR gate altering the detector identification pulse by producing multiple pulses instead of a single pulse.

For detector 0, an averaged original pulse and averaged multiplexed pulse are shown in figure 15. Averaged pulses for the other detectors were also found to be similar. TMUX 1101 has a typical on-resistance of 3.7 Ω for $V_DD = 3.3$V at 25°C [19]. As a result, the amplitude of the switch output is attenuated. The amplitude of the switch output can be adjusted by adjusting the gain of the non-inverting amplifier of the corresponding attenuated signal amplifier and copier (ASAC) unit.

The linearity of the proposed multiplexing method was examined in terms of the charge collected under the original pulse and the multiplexed pulse. The DT2740 ADC has an input impedance of 50 Ω, and it records any signal as a voltage waveform. The total charge ($Q$) collected under any anode pulse is given by $Q = \frac{1}{R} \cdot \int_0^t v(t')dt'$, where, $R$ is the input impedance of the ADC, $v(t')$ is the amplitude of the voltage pulse at any time $t'$, and $t$ is the pulse duration in units of time. The sampling interval was 8 ns and the recorded pulse amplitude was in the scale of mV. Therefore, trapezoidal integration over time yielded charge collection in scale of pC. Figure 16 shows the linear relationship between the charge collected under the original pulse and the multiplexed pulse for detector 0. The deviation of the multiplexed pulse from the strong positive correlation with the original pulse, in terms of the collected charge, was also measured. Figure 17 shows distributions of deviation for different collected charge ranges, and figure 18 shows the mean deviations for the same collected charge ranges for detector 0. For all four detectors, the charge collected under the multiplexed pulse was found to be linearly proportional to the charge collected under the original pulse. Also, no significant variation in the mean deviation of the multiplexed pulse for different collected charge ranges was observed.

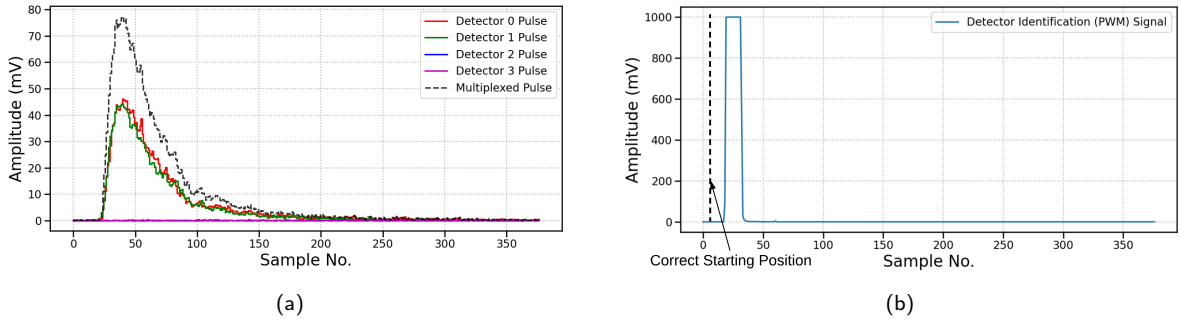

**Figure 14:** Demonstration of the XOR gate altering the detector identification pulse by shifting the pulse to the right relative to its usual position in a digitized record.

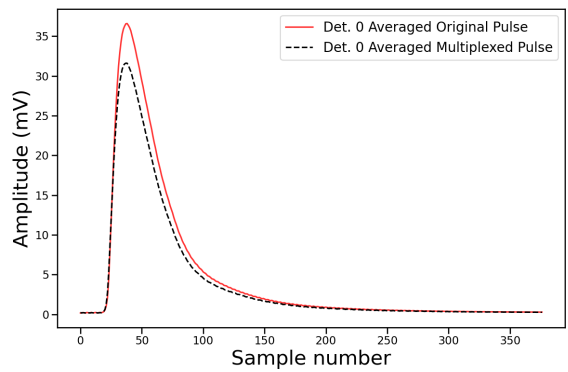

**Figure 15:** Averaged original pulse and averaged multiplexed pulse for detector 0.

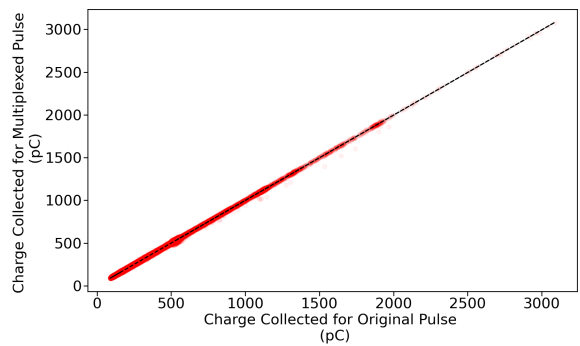

**Figure 16:** Linear relationship between the charge collected under the original pulse and the multiplexed pulse for detector 0.

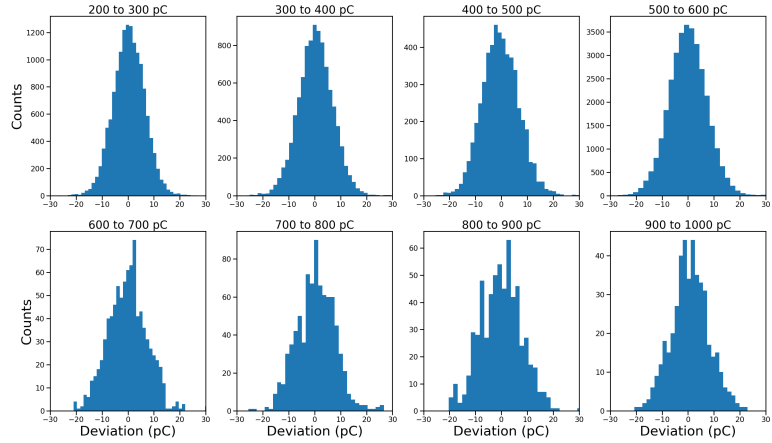

**Figure 17:** Distributions of charge deviation of the multiplexed pulse from the strong positive correlation with the original pulse for detector 0. Distributions are presented for 8 different collected charge ranges.

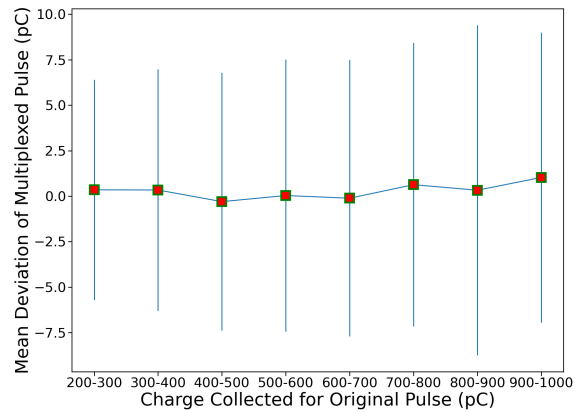

**Figure 18:** Mean charge deviation of the multiplexed pulse for different collected charge ranges of the original pulse for detector 0.

## 5.2. Energy Resolution Measurements

In order to evaluate the degradation in energy resolution caused by the multiplexer, calibrations were performed using the $^{137}$Cs photopeak to construct energy spectra in terms of keV. The energy spectrum for a $^{137}$Cs source, obtained using detector 0 with an analog pulse processing system is shown in figure 19. The energy resolution of detector 0 at the 662 keV peak was found to be 7.113% ± 0.003%. Here, 0.003% is the standard deviation of the energy resolution obtained by propagating uncertainty in the Gaussian fit to the photopeak. It implicitly includes uncertainty introduced by baseline subtraction, because charge was integrated from baseline-subtracted pulses. For a four-channel setup, the energy spectrum for the same source and detector, obtained by digitizing the original pulses, is shown in figure 20. The estimated energy resolution of detector 0 at the 662 keV peak was found to be 7.083% ± 0.036%[4]. The energy spectrum for the same source and detector, obtained using pulses output by the four-channel multiplexer, is shown in figure 21. For the multiplexed pulses, the estimated energy resolution of detector 0 at the 662 keV peak was found to be 7.209% ± 0.037%. Therefore, a small degradation of energy resolution of about 0.13% was observed for detector 0 using a four-channel multiplexer. For both original and multiplexed pulses, energy calibrations were performed separately as the values of the measured charge for the same photon were different due to slight differences in PMT gain. Energy resolution measurements were also performed using one-channel, two-channel, and three-channel setups. The two-channel setup used detectors 0 and 1 and the three-channel setup used detectors 0, 1, and 2. Four one-channel setups were built using each detector. Degradation of energy resolution as a function of the number of multiplexer channels was studied to observe the effect of insertion of off-state leakage current from the inactive switches. TMUX1101 [19] has a rated off-state leakage current of 3 pA which can degrade the energy resolution by adding up at the fan-in circuit as the number of channels increases. Nevertheless, in all cases, a similar small loss of resolution was observed for all four multiplexer channels. For four-channel, three-channel, two-channel, and one-channel setups, loss of energy resolution of about 0.12% to 0.16%, 0.07% to 0.12%, 0.11% to 0.17%, and 0.05% to 0.16% were observed, respectively. The standard deviation of the fitted energy resolution was below 0.04% for all cases.

---

[4]The number of events (∼100,000 per detector) recorded with the digitizer was significantly smaller than the number events(∼12 million per detector) recorded with the analog measurement system, which resulted in ∼10 times higher uncertainty then the reference uncertainty (0.036% vs 0.003%).

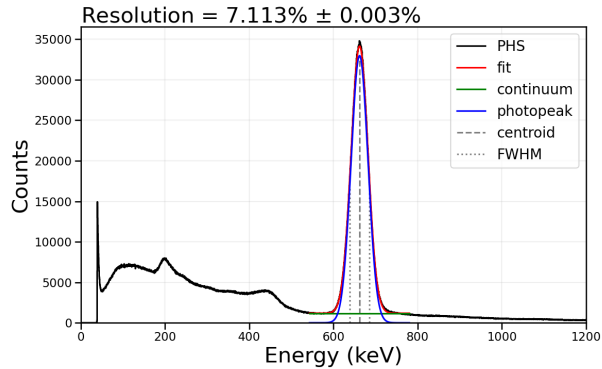

Figure 19: The energy spectrum for a $^{137}$Cs source obtained with detector 0 and an analog pulse processing system.

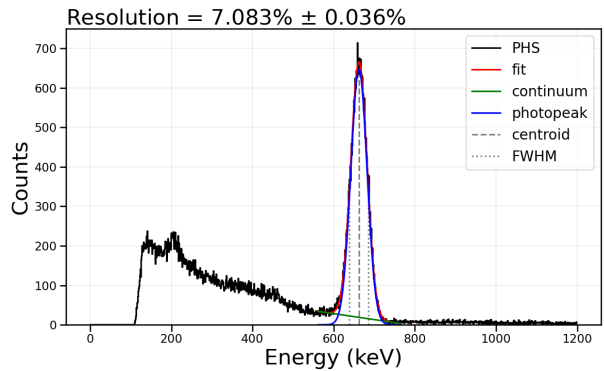

Figure 20: The energy spectrum for a $^{137}$Cs source obtained with the original pulse from detector 0 and the four-channel multiplexer setup.

The x-ray peak evident on the left side of figure 19 was suppressed by the trigger level of the multiplexer's front-end leading edge discriminator.

Average energy resolution for each detector for all four setups was also estimated using the bootstrapping method, also referred to as sampling with replacement [24]. It was used to verify that the sample data set was unbiased. In bootstrapping method, 10 sub-samples were constructed by taking random events from the known sample, with replacement. Each sub-sample contained 10,000 events. Energy resolution for each sub-sample was estimated and the average energy resolution for each detector was estimated from the mean of the sub-sampled datasets. For four-channel, three-channel, two-channel, and one-channel setups, average loss of energy resolution of about 0.08% to 0.15%, 0.04% to 0.12%, 0.11% to 0.14%, and 0.06% to 0.12% were observed, respectively. For almost all cases, the standard deviation of the average energy resolution was below 0.04%. By comparing the energy resolution obtained from the known sample and the bootstrapped sub-samples, no bias was observed in the sample data set.

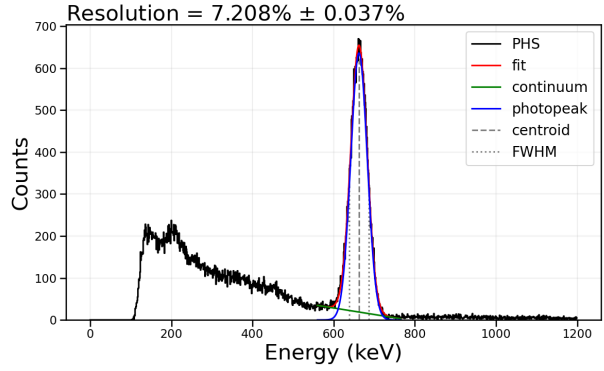

Figure 21: The energy spectrum for a $^{137}$Cs source obtained with the multiplexed pulse from detector 0 and the four-channel multiplexer setup.

### 5.3. Timing Resolution Measurements

In order to evaluate degradation in timing resolution caused by the multiplexer, timing resolution measurements were performed using the setup described in section 4.2. The distribution of difference in the time-of-arrival (TOA) of coincident 511 keV annihilation photons from a $^{22}$Na source is shown in figure 22. For the time pick-off, constant-fraction discrimination (CFD) was applied to the original pulses and the multiplexed pulses. The constant fraction value was set to 0.4, and the CFD delay was set to 24 ns. The estimated FWHM of the distribution for the original pulse and the multiplexed pulse were found to be 10.806 ns ± 0.005 ns and 11.027 ns ± 0.007 ns, respectively. A small increase of the FWHM of about 221 ps for the multiplexer was observed; this is substantially smaller than 8 ns sampling interval of the digitizer and the ∼10 ns timing resolution intrinsic to NaI.

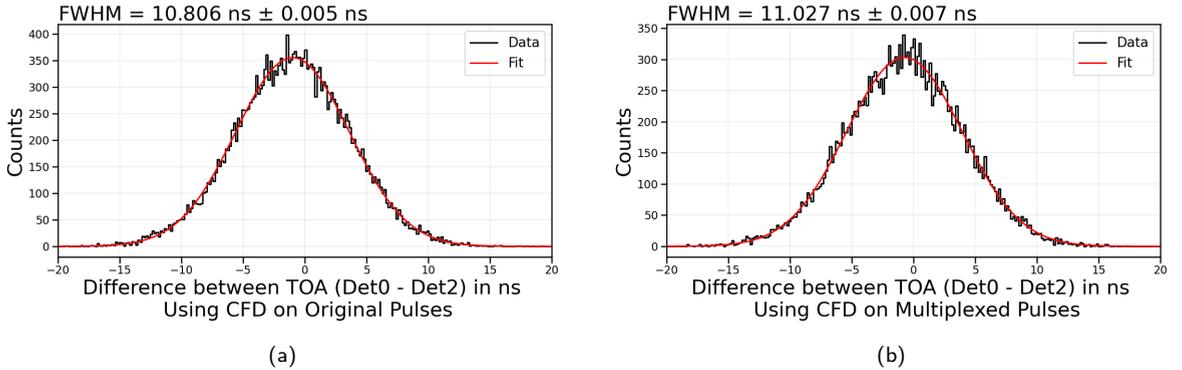

Figure 22: The distribution of difference in the time-of-arrival (TOA) of the 511 keV coincidence gamma rays at both detectors by applying CFD to (a) the original pulses and (b) the multiplexed pulses. A small loss of timing resolution, ∼221 ps, was observed using the multiplexed pulses.

### 5.4. Pulse Shape Measurements

A series of measurements were also conducted to evaluate changes in anode pulse shape caused by the multiplexer. The rise-time constant ($\tau_r$) and fall-time constant ($\tau_f$) for each anode pulse were estimated by fitting the recorded pulses using equation 1 [25].

$$V(t) = u(t) \cdot A \cdot \frac{\lambda}{\lambda - \theta} \cdot \left[ e^{-\theta(t - t_0)} - e^{-\lambda(t - t_0)} \right] \tag{1}$$

Here, $u(t)$ is the unit step function, and

$$A = \frac{Q}{C}$$
$$\tau_r = \frac{1}{\lambda - \theta}$$
$$\tau_f = \frac{1}{\theta}$$

where $Q$ is the total charge collected under the pulse, $C$ is the PMT's equivalent capacitance, $\lambda$ is the scintillator decay constant, and $\theta$ is the reciprocal of the PMT's RC time constant. A sample fitted pulse is shown in figure 23.

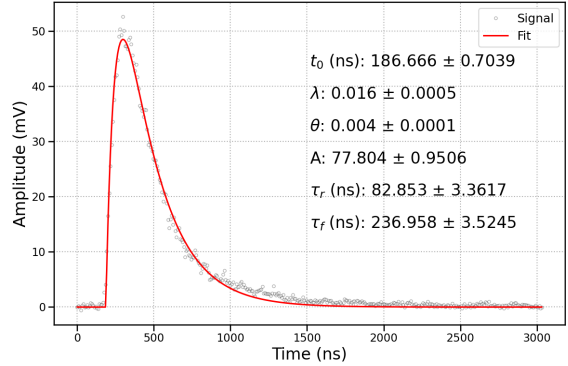

**Figure 23:** A sample fitted anode pulse.

The distribution of rise-time constant ($\tau_r$) and fall-time constant ($\tau_f$) for both original pulses and multiplexed pulses from detector 0 are shown in figure 24 and figure 25, respectively. For other detectors, similar distributions were observed. After comparing the distributions for original pulses and multiplexed pulses, no significant difference was observed.

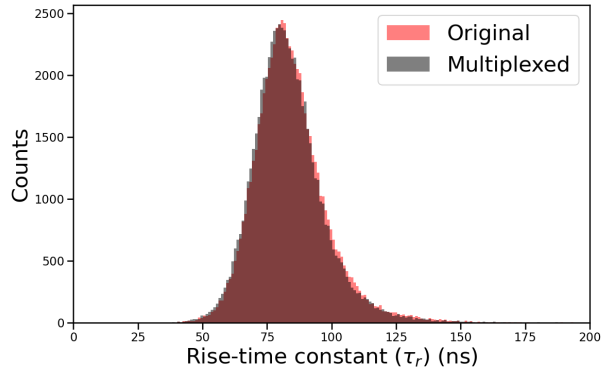

**Figure 24:** Distribution of rise-time constant ($\tau_r$) for detector 0.

The linear relationship between the original pulse's time constants (both $\tau_r$ and $\tau_f$) and the multiplexed pulse's time constants for detector 0 is shown in figure 26 and figure 27. For all four detectors, the time constants of the multiplexed pulse were found to be linearly proportional to the time constants of the original pulse. Events deviating from proportionality were found to have uncertainty in the fit parameters which can be observed from the large error bars for those events. In most cases, the large value of uncertainty in estimating $\tau_r$ or $\tau_f$ occured for very low amplitude pulses (typically around 10 mV) where the pulse amplitude becomes comparable to the noise amplitude.

A Hardware/Firmware-Based Switching Gate Multiplexing Method for Pulse Mode Radiation Detectors

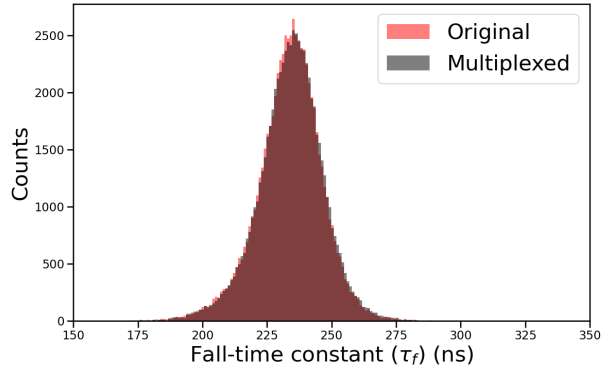

**Figure 25:** Distribution of fall-time constant ($\tau_f$) for detector 0.

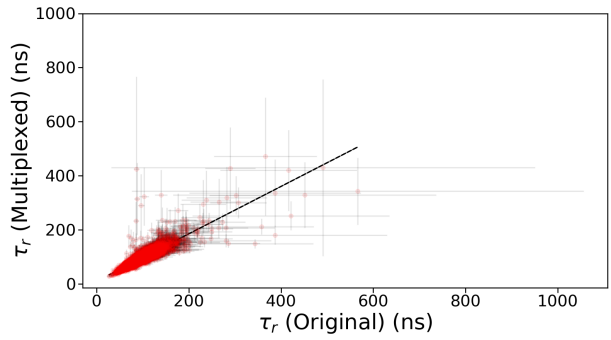

**Figure 26:** The linear relationship between the rise-time constant ($\tau_r$) of the original pulse and the multiplexed pulse for detector 0.

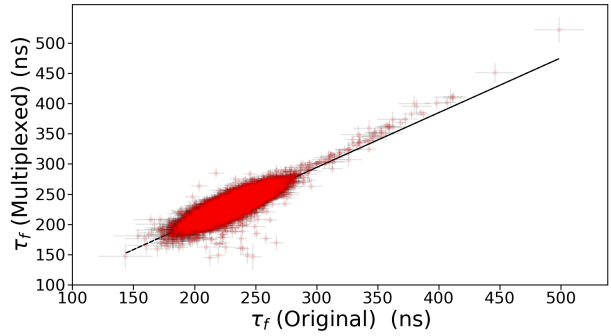

**Figure 27:** The linear relationship between the fall-time constant ($\tau_f$) of the original pulse and the multiplexed pulse for detector 0.

The original and multiplexed pulses were also compared in terms of the mean absolute relative deviation (MARD) of both $\tau_r$ and $\tau_f$. MARDs were measured using equation 2.

$$MARD = \frac{1}{N} \sum_i \left| \frac{\tau_i}{\bar{\tau}} - 1 \right| \qquad (2)$$

$$\sigma_{MARD} = \frac{1}{N\bar{\tau}} \sqrt{\left( \sigma_{\tau_1} + \sigma_{\tau_2} + ... + \sigma_{\tau_N} \right)} \qquad (3)$$

where $\tau_i$ is the time constant for the $i^{th}$ event and $\bar{\tau}$ is the mean time constant.

MARDs for all detectors of a four-channel multiplexer were measured. For detector 0, the MARD of $\tau_r$ for the original and multiplexed pulses were found to be $0.1942 \pm 0.0006$ and $0.1929 \pm 0.0006$, respectively. The MARD of $\tau_f$ for the original and multiplexed pulses were found to be $0.05660 \pm 0.00009$ and $0.0569 \pm 0.0001$, respectively. Similar results were also observed for all other detectors, too. Like the previous two analyses, no significant difference between the original pulse and the multiplexed pulse was observed in terms of MARD.

## 6. Conclusion and Future Work

The proposed hardware/firmware-based switching gate multiplexing technique was demonstrated by multiplexing four NaI(Tl) inorganic scintillator detectors. The average of original pulses and multiplexed pulses were measured and were found to be consistent in shape. The multiplexer's ability to identify and recover the detector pulse coming from the first active detector when multiple pulses from different detectors coincide was also demonstrated. The performance of the multiplexing technique was evaluated by comparing its timing and energy resolution and pulse shape to that obtained with un-multiplexed signals. The energy spectrum of a $^{137}$Cs source was measured by using an analog pulse processing system and different multiplexing setups consisting of different number of detectors. The multiplexing method did not exhibit any substantial change in energy resolution degradation when the number of detectors was increased. The timing resolution was estimated using coincidence measurements with a $^{22}$Na source. A small loss of timing resolution, about 221 ps, was observed. This degradation is substantially smaller than 8 ns sampling interval of the DT2740 digitizer. The multiplexer's ability to retain the original pulse shape was evaluated by comparing the distribution and mean absolute relative deviation (MARD) of time constants for the multiplexed pulses to that obtained for the original pulses. From these comparisons, no significant change in pulse shape was observed. Moreover, time constants for the multiplexed pulse were linear with that obtained for the original pulse. The multiplexer retains the original pulse shape by passing the detector pulse to the fan-in circuit through an analog switch. This feature makes the performance of the multiplexer dependent on the performance parameters of the op-amps and analog switch ICs, such as high bandwidth, low input noise, low charge injection etc. As a result, the performance of this multiplexer can be improved in the future by using superior op-amps and analog switch ICs available at that time. These features make the hardware/firmware-based switching gate multiplexing method a promising solution to be applied in large detection networks.

The proposed multiplexing method is designed to multiplex many detector signals into two readout channels; one digitizes the signal that contains multiplexed detector pulse, and the other digitizes the signal that contains an identification pulse. As a result, a multiplexing ratio of N:2 can be achieved for N detectors. A multiplexing ratio of N:1 can be achieved by tagging the identification pulse at the tail of multiplexed pulse [14]. This can be done very easily by delaying the identification pulse by the FPGA and passing it to the summing amplifier or the fan-in circuit instead of passing it to an ADC channel. However, due to the addition of this active signal to the input of the fan-in circuit, noise from that signal would be added to the multiplexed output. Therefore, the performance of the N-to-1 multiplexer needs to be evaluated and compared to that of the N-to-2 multiplexer. This work will be performed in the future.